# Extending Electrical Resistivity Measurements in Micro-scratching of Silicon to Determine Thermal Conductivity of the Metallic Phase Si-II

**Hisham A. Abdel-Aal, Ysai Reyes**
Department of General Engineering,
University of Wisconsin-Platteville,
1 University Plaza, Platteville,
Wisconsin 53818-3099
abdelaah@uwplatt.edu

**John A. Patten,**
Department of Manufacturing Engineering,
Western Michigan University,
C-245 Parkview Campus, Kalamazoo,
Michigan 49008-5336

**Lei Dong**
Department of Mechanical Engineering and Engineering Science
The University of North Carolina at Charlotte
9201 University City Blvd.
Charlotte, NC 28223-0001

## ABSTRACT

*The implementation of many Micro Electronic Mechanical Systems (MEMS) involves several silicon machining processes. These processes introduce structural damage in the material. The trend in removing this damage is to apply finishing techniques such as grinding or lapping, etc. Since silicon is a brittle material, the removal of material is expected to take place through the propagation of fracture. This renders the machining process more difficult to control. Material removal in the ductile regime, which is more attractive for surface quality purposes, was theorized to be possible in the late eighties. Since Ductile Regime Machining (DRM) is favorable, due to the higher quality of the resulting surface, it is required to understand the mechanism by which silicon behaves as a ductile material. This hinges on understanding the energetics of silicon metallization under pressure. Fundamental to such endeavor is the to develop a parametric thermal model of the DRM process. Such a model is not yet possible since the metallic phases of silicon aren't readily amenable to thermal characterization through direct measurements. This being the case, one has to rely on processing indirect measurement data to deduce refined estimates of the thermal transport properties of pressurized Silicon. One of the most popular, perhaps more established, measured quantities is the electrical resistivity. In fact, many researchers monitor the variation in the electrical resistivity of silicon during indentation and consider its' change as an indicator of the formation of a metallic silicon phase under the indenter. This metallic state is advantageous from a thermal transport property characterization prospective. Since the material under the indenter is metallic the electrical resistivity and the thermal conductivity of the transformed phase should obey the Wiedman-Franz- Lorenz law. That is, both these properties are interrelated and can be*

*estimated if either of them is known along with the temperature at the specific point with any certainty. This paper, therefore, describes a procedure by which the average thermal conductivity of the metallic phase of silicon is extracted from electrical resistivity measurements. The procedure is based on teaming a temperature evaluation code to the resistivity measurements thus allowing the extraction of the conductivity as a function of temperature.*

# Introduction

The mechano-physical characterization of Si under the influence of indentation pressure and in sliding contact has been the focus of numerous investigations. This is not surprising since the quality of a machined silicon device is directly related to the quality of the surface of the wafer after slicing it from a bulk Si ingot.

In order to produce near perfect surface quality of Si wafers, a sequence of manufacturing processes is required. These vary between lapping and grinding or a combination of both. Each of these processes introduce a ceratin amount of lattice damage in the near surface wafer regions. This damage is typically removed by chemical etching which affects the flatness of the wafer surface. To avoid the degradational effects of etching high quality grinding is normally applied to the wafer followed by final polishing. In such a manner, not only the down effects of etching are avoided, but also the potential for polishing induced surface damage is reduced. Depending on the final device application, further wafer manufacturing steps may follow wafer polishing. The details of any sequence of processes that the wafer undergoes depends on the final device. However, regardless of the details, the goal of the manufacturing processes remains to ensure a high surface quality of the wafer front side. This requires consistent control and improvement of sequential processing steps [1].

Achieving a superior level of control and consistency is not an easy task in machining of silicon. This is because silicon is a brittle material. As such material removal is expected to be mainly due to propagation of fracture. Such regime of machining is not necessarily conducive to the level of control of surface quality require for the production of many MEMS/NEMS. In the late eighties of the 20$^{th}$ century, the possibility of machining silicon in the ductile regime was proposed by Blake and Scattergood[2]. Following this pioneering work several researchers have endeavored to explain the origins of silicon ductility in machining [3-5].

The removal of silicon in the ductile regime is technologically attractive since it offers more control of the resulting surface. This would lead to a near perfect wafer surface as it allows to suppress brittle fracture, decrease surface damage and improve quality of machining [2,3, 6-9]. Being that advantageous, the origins of ductility in silicon and semiconductors in general are a subject of continuous investigations. These investigations seek the origins of the ductile behavior of Si and the physics of its behavior under indentation [10-13].

Few published work, if any, have sought to characterize the engineering properties of the ductile silicon phase. There are many reasons for such state of affairs among which is the lack of understanding of the genesis of silicon ductility combined to the diverse theories that attempts to explain such ductility. These theories orbit four main schools of thought. These vary between: classical dislocation [10],microfrature [11], thermal melting [12], and phase transformation [13]. Given this diversity, it might be argued that the characterization of the engineering properties of the ductile phase is a premature step. However, for the practitioner it is of essential importance to have an estimate of such properties. This is because the necessity of the parametric study of DRM in order to optimize wafer production sequences.

This paper is a step in this direction. In this work we attempt to extend the validity of electrical resistivity measurements of silicon under indentation pressure. The goal of such extension is to estimate the thermal conductivity of the transformed silicon layer which lies directly under the indenter.

## 2. Behavior of Si during scratching and ultra-precision cutting

The two operations involving dynamical contact between a single asperity and the specimen are scratching – that is, sliding of the hard indenter/stylus across the sample surface, and ultra precision cutting – that is, removal of workpiece material using an inclined tool with a large edge radius relative to the depth of cut, or a sharp tool with a negative rake angle. In ultra precision cutting, the clockwise rotation of the resultant force leads to the formation of chips ahead of the tool rather than debris alongside the grooves, typical for indentation sliding. In both cases, however, the material may be removed in a ductile regime that involves plastic flow (severely sheared chips or debris), without introducing fracture damage (propagation and intersection of cracks) into the machined surface.

A comprehensive study of Si response to scratching has been performed by Gogotsi et al. [14]. A single point diamond turning machine was used to make grooves in (111) silicon wafers at room temperature. Both sharp (Vickers) and blunt (Rockwell) diamond indenters were used for scratching. The depth of cut was increased gradually to monitor the transition from ductile to brittle regime of material removal. Post-scratching phase and stress analysis was done by means of Raman micro-spectroscopy, and the grooves morphology was characterized by scanning electron microscopy, atomic force microscopy, and optical profilometry.  The typical spectra taken from the areas within the grooves showed the presence of the r8 (Si-XII) and bc8 (Si-III) polymorphs of silicon. The debris/chips produced Raman spectra with a "doublepeak" feature, indicative of the hexagonal Si-IV phase or nanocrystalline material. The beginning portions of the grooves (at small depths of cut) are mostly amorphous. This perfectly correlated with the observations of static contact loading, when rapid unloading from the metallic Si- II phase leads to the formation of amorphous silicon, whereas slow unloading produced the mixture of the Si-III and Si-XII phases.   Silicon response to scratching in the ductile regime is thus envisaged in the following fashion, figure 1,:  Highly localized stresses underneath the tool lead to the formation of the metallic Si-II phase, which deforms by plastic flow and subsequently transforms into a-Si or a mixture of Si-III and Si-XII behind the tool. These reverse phase transformations, accompanied by a ~10% volume increase, are at least partly responsible for the complex groove morphology after the load is removed.

The findings of Gogosti and coworkers regarding the metallic state of the silicon layer under the tip of a machining tool or a scratching stylus are consistent with the findings of many other workers who studied the behavior of Silicon under indentation. So that, in essence, the behavior of silicon under dynamic loading or under static indentation is the same. The metallic nature of the transformed silicon layer may facilitate the prediction of the thermal properties directly under the tool/indenter.  This is achieved by applying the so called Wiedman-Franz-Lorenz law.

## 3. The Wiedemann-Franz-Lorenz law

The electron fluid in a metal is an excellent conductor of electrical charge. It is also an excellent conductor of heat. In an insulator phonons (quantized lattice vibration) carry all the heat current. By contrast, in familiar metals, the electron fluid conducts nearly the entire heat current ( the phonon current still exists but constitutes an extremely small percentage of the total heat current). Although the electrical and thermal currents carry distinct quantities ( charge and entropy) it is fruitful to compare their magnitudes. The earliest comparison of these two analogous quantities, the thermal conductivity K and the electrical conductivity σ was made by Wiedemann and Franz in 1853.

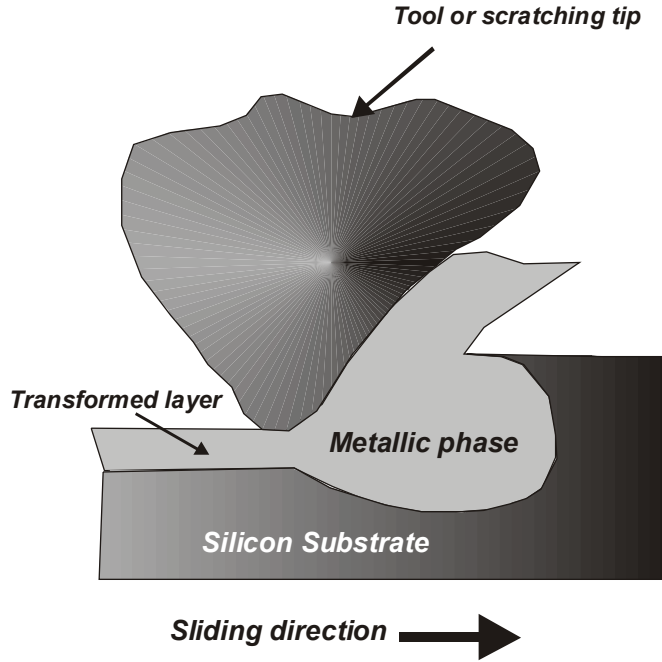

*Figure 1:   Response of Bulk silicon to Scratching. Highly localized stresses underneath the tool lead to the formation of the metallic Si-II phase, which deforms by plastic flow and subsequently transforms into a-Si or a mixture of Si-III and Si-XII behind the tool*

Wiedemann and Franz used a classical approach based upon the fact that the heat and electrical transport both involve the free electrons in the metal. The thermal conductivity increases with the average particle velocity since that increases the forward transport of energy. However, the electrical conductivity decreases with particle velocity increases because of the collisions divert the electrons from forward transport of charge. This means that the ratio of thermal to electrical conductivity depends upon the average velocity squared, which is proportional to the kinetic temperature. Based on their classical mechanics approach Wiedemann and Franz deduced that the ratio of thermal to electrical conductivity is given by:

$$\frac{K}{\sigma} = 3 \left(\frac{K_B}{e}\right)^2 T = 2.23*10^{-8} \; W\Omega K^{-2} \tag{1}$$

Where $K_B$ is the Boltzman's constant and $T$ is the absolute temperature. This remarkable result predicts that the Wiedemann Franz ratio doesn't depend on either the relaxation time or electron density and is thus independent of any property of the metal. In other words, the ratio $K/\sigma T$ should have the same value at all temperatures for all metals. The advent of quantum mechanics prompted recalculation of the Wiedemann-Franz ratio. It was found that the quantum result for two components of $K/\sigma T$ are totally different from their classical values. However, due to a fortunate cancellation of factors, the rather incorrect classical result, given by equation(1) is surprisingly similar to the correct quantum mechanics-based result i.e.,

$$\frac{K}{\sigma} = \frac{\pi^2}{3}\left(\frac{K_B}{e}\right)^2 T = 2.45*10^{-8} \; W\Omega K^{-2} \qquad (2)$$

The value calculated by the methods of quantum mechanics is called the Lorenz number and is in close agreement with the value predicted by equation (1).

In a similar manner, the quantum mechanics-based Lorenz number for semi-conductors is given by [15]:

$$\frac{K}{\sigma} = \frac{8}{\pi}\left(\frac{K_B}{e}\right)^2 T = 1.856*10^{-8} \; W\Omega \, °K^{-2} \qquad (3)$$

The behavior of normal metals and alloys is accurately predicted by the Wiedmann-Franz-Lorenz law. Semi-conductors, however, show greater deviation from this law. This is attributed to the number of non-degenerate electrons involved and the major role relegated to the lattice in thermal conduction in these materials. As such, considering the metallic nature of the transformed silicon, equation (2) would be more suitable to apply.

The Wiedman Franz law involves three quantities: the absolute temperature, the electrical conductivity, and the thermal conductivity. Thus, to apply equation (2) to silicon, two of the three involved quantities have to be readily available. These are, the absolute temperature of the silicon layer, the electrical conductivity (or alternatively the electrical resistivity). These two quantities may be determined by direct measurement or in case of the temperature by applying the appropriate solution of the heat equation. This later approach is adopted in this work.

### 4. Temperature rise due to a moving heat source

The problem of the temperature rise due to a moving heat source have been exhaustively investigated. There are several direct analytical solutions for the heat equation both for the steady and the transient conduction.

The distinction between the steady state and the transient state depends on the Peklette number which is given by:

$$Pe = \frac{Va}{D} \qquad (4)$$

where V is the speed of the heat source, *a* is the radius of the source and *D* is the thermal diffusivity of the body. When the Peklette number is less than unity the source is considered quasi-stationary and steady state conditions are assumed to prevail. On the other hand, if the Peklette is higher than 5 then the source is considered to be fast moving and transient conditions are supposed to prevail. The solution of the heat equation for the quasi-stationary source is due to Jaeger [15] and is given by:

$$\theta = \frac{Q}{\pi K} \int_{-a}^{a} Exp\left\{\frac{V(x-\bar{x})}{2D}\right\} \cdot K_o\left\{\frac{V(x-\bar{x})}{2D}\right\} d\bar{x} \qquad (5)$$

Equation (5) may be integrated in a close form as follows:
let

$$U = \frac{V(x-\bar{x})}{2D}, \qquad X = \frac{Vx}{2D} \qquad (6\text{-a})$$

then

$$dU = -\frac{Vx}{2D} \qquad (6\text{-b})$$

therefore,

$$\theta = \frac{Qa}{2\pi KR} \int_{X-2R}^{X+2R} Exp\{U\} \cdot K_o |U| \, dU \qquad (7)$$

Equation (7) can be integrated to:

$$\theta = \frac{Q\,a}{2\pi K R} [X+2R]\,Exp\,[X+2R]$$

$$\left[ K_o[X+2R] - \frac{|X+2R|}{\{X+2R\}} K_1|X+2R| \right]$$

$$- [X-2R]\,Exp\,[X-2R]$$

$$\left[ K_o[X+2R] - \frac{|X-2R|}{\{X-2R\}} K_1|X-2R| \right]$$

(8)

Where $R$ is given by:

$$R = \frac{V_{slid} * a}{D}$$

it is noted here that $R$ is a figure of merit that is analogous to the Peklette number alternatively called the *Jaeger* number.

## 4. Measurement of the electrical resistivity

Electrical resistivity and Electrical conductivity are two reciprocal quantities. So that it suffices to measure either quantity to deduce the other. In this work, data pertaining to the electrical resistivity of a silicon wafer were processed using the Wiedemann-Franz-Lorenz law. Experimental setup and measurement procedures are detailed elsewhere [16]. Table 1 presents a summary of the specifications of the apparatus and materials used.

**Table-1** Summary of the experimental conditions and specifications of the apparatus used in measuring the electrical resistivity

| | | |
|---|---|---|
| *Wafer* | Silicon (Si) | 111 –type, 1-10 ohm - cm |
| | Gold (Au) | Thickness: 0.3 µm<br>Width: 100 µm |
| *stylus* | Powered by | Surface profilometer (Surfcom 110 B, Brown and Sharpe Co.) |
| | Head | Diamond, 9 µm radius (estimated from SEM image and scratch geometry based upon AFM analysis) |
| | Speed | Automatic : 0.305 mm/sec *forward*<br>1.5 mm/sec *backward*<br>Manual: optional (slower than automatic) |

| | |
|---|---|
| *Load* | 3g, 5g, 10g |
| *Contact Probes* | Rucker& Kolls (model number 329-6) |
| *Constant Current Source* | 1 mA (approximately) |
| *Voltmeter* | Labview (National Instruments Max Voltage: 10 V |
| *Cleaning* | Flux remover (Flux- OFFROSIN, Chemetronics) |

## 4.1 Experimental Procedure

Two scratching methods were employed. The first is *Automatic Scratching* whereas the second is *Manual Scratching*. Both methods utilize a stylus of a surface profilometer (Surfcom 110B, Brown & Sharp Co.). Figure 2 -a depicts a typical spherically capped diamond stylus used in scratching. Whereas, figure 2-b depicts a typical scratching specimen with Gold terminals deposited on the surface of a silicon wafer.

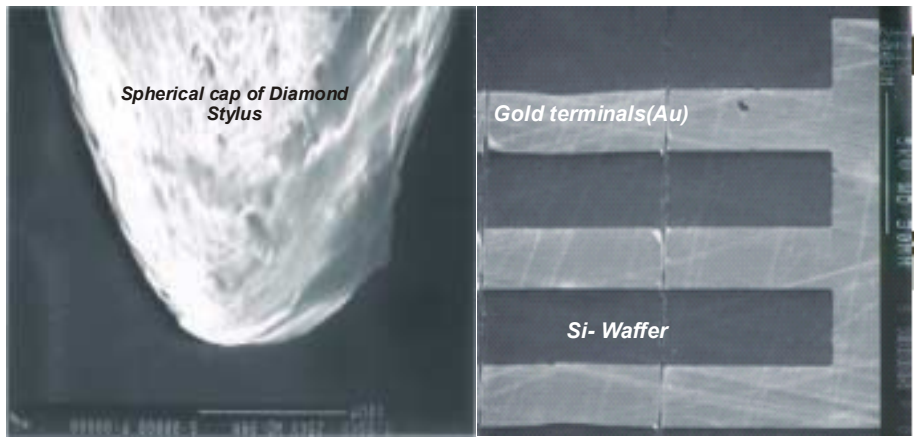

*Figure 2: Typical samples used in the scratching experiments*
  *a- a spherically capped diamond stylus*
  *b- gold terminals deposited on the surface of a silicon wafer*

The experimental procedure was as follows:
(a) A constant current (1mA) flows through a gold contact pad on a Silicon wafer and the voltage across the gold pad is measured
(b) A loaded stylus powered by the profilometer moves across the gold pad, perpendicular to the current flow, and produces a scratch
(c) After several scratches in the same groove, the stylus removes the gold and separates the gold pad into two sections. Then the Silicon beneath the gold is exposed and the stylus penetrates the Silicon surface in subsequent scratches. Hence the first several scratches are in preparation for measuring. (Note: The Silicon on either side of the gold pads also scratched.
(d) When the stylus moves in and out of the gap (created by the above scratches) between the separated gold pads, the voltage changes, i.e. the resistance changes (since V=I*R, I: constant. The high voltage (resistance) corresponds to the semiconducting Silicon, while the low voltage (resistance) corresponds to metallic Silicon.
(e) The voltage is measured according to time instead of the stylus position, the stylus position is in proportion to time when the stylus moves at constant speed.

### 4.2. Parameters

The electrical resistance measurement is made with the following parameters.

(a) *Load (Pressure)*

The stylus is preloaded with one of three weights, light (3g), middle (5g) and heavy (l0g), to observe the influence of the pressure on the Silicon, and the extent of the high pressure zone.

(b) *Stylus Speed - Automatic and Manual scratching*

The stylus scratches the Silicon at different speeds to observe the influence of the cutting speed as follows:

(1) *Automatic scratching* (Stylus speed: 0.305mm/sec

The profilometer is operated as normal, except that its stylus is preloaded with a weight. The voltage measurements across the gap are made during scratching.

(2) *Manual scratching* (Stylus speed: variable, slower than Automatic)

After Automatic scratching, the stylus is moved manually, with slower speed than in Automatic mode, on the same groove made by previous Automatic scratching, and the voltage during Manual scratching is measured.

### 5. Results and concluding remarks

The results presented herein pertain to the analysis of voltages obtained for the 5g manual scratching experiments. Figure 3 is a schematic of the equivalent resistance circuits during scratching.

### 5.1 analysis procedure

According to figure 3, the resistance of the silicon metallic phase $Si_m$ is given by:

$$R_{Si_m} = \frac{R_{si}[R - R_{Au}]}{R_{si} - [R - R_{Au}]} \qquad (9)$$

where, $R_{si\,m}$ is the resistance of the metallic phase in Ohms, R is the general resistance of the specimen with the stylus loaded, $R_{Au}$ is the resistance of the gold terminals and $R_{si}$ is the resistance of the untransformed (covalent) silicon.

The resistance $R_{si\,m}$ which is determined from equation (9) is then divided by the depth of the scratch profile, $\lambda$, as determined from surface profilometry of the scratched sample, and then multiplied by the area of contact between the diamond stylus and the silicon wafer. This later quantity is determined from Hertzian contact theory. As, such the electrical resistivity of the metallic phase is given by:

$$\rho = \frac{R_{Si_m} \lambda}{A}$$

Figure 2 depicts the variation in the resistivity of the metallic phase with the strength of the applied current. It is noted that the resistivity initially increases with increasing the strength of the current then experiences a drop as the applied current increases. The resistivity resulting from the application of equation (9) is then used to deduce the thermal conductivity of the metallic phase by the simultaneous application of the Wiedemann-Franz-Lorenz law and the temperature expression given by equation (8) in an iterative procedure. This procedure can be proved to converge rapidly (order of 10-15 iterations). Figure 4 depicts the variation in the relative error in computing the thermal conductivity of the metallic phase with the number of iterations.

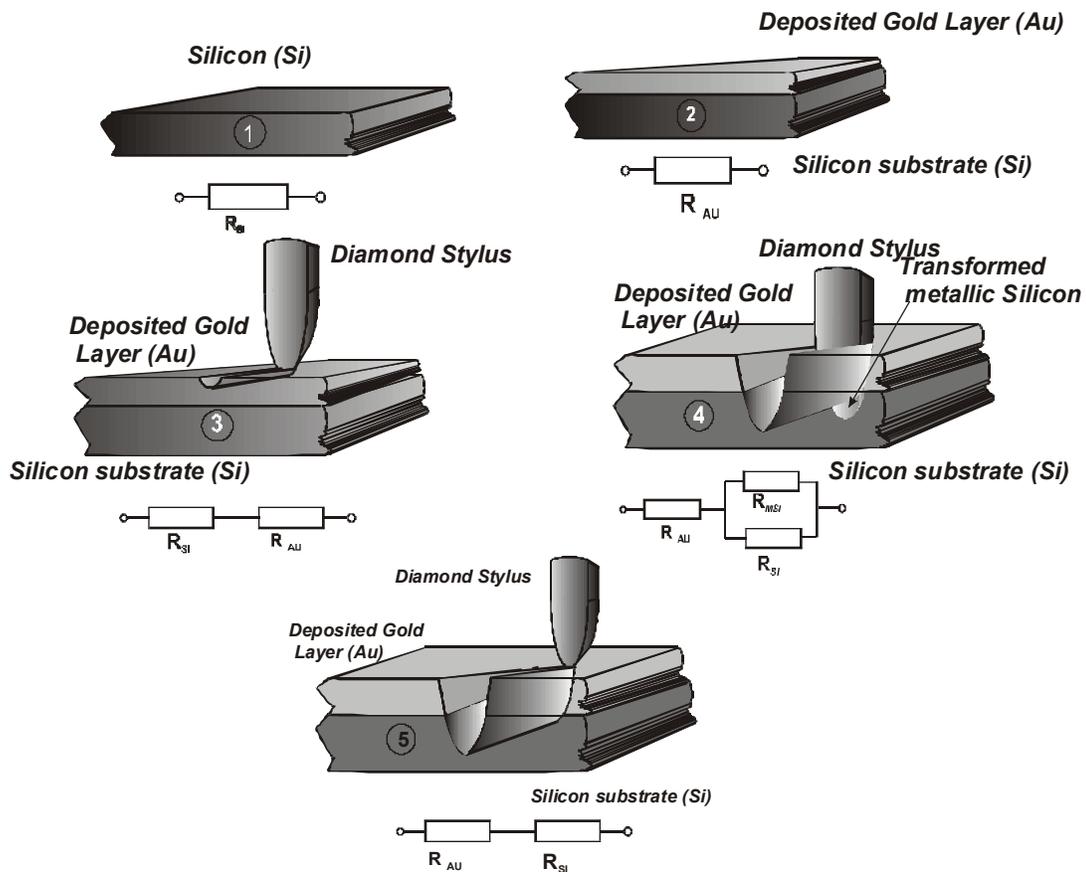

Figure 3: Schematic of the equivalent electrical resistance during the different stages of scratching a silicon wafer
    1- **Silicon wafer before deposition of Gold terminals**
    2- **after deposition of a Gold (AU) layer**
    3- **Diamond stylus scratching the Gold layer**
    4- **Diamond stylus scratching both gold and transformed silicon**

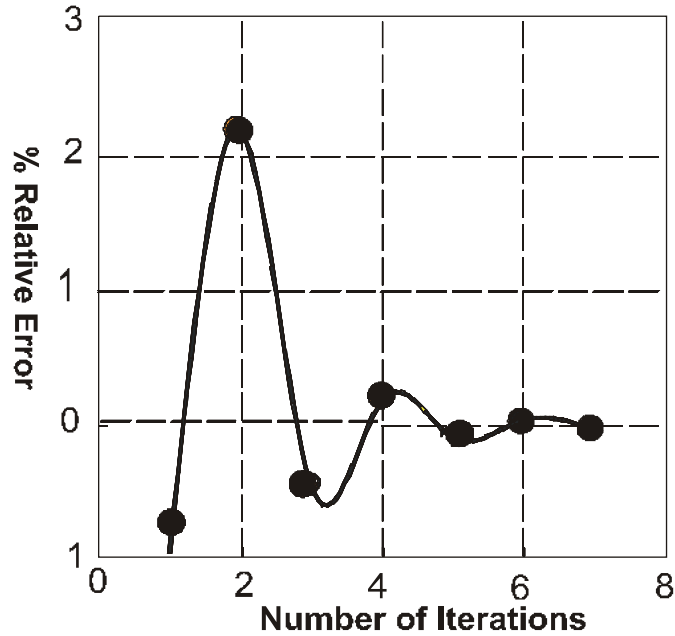

*Figure 4     Propagation of the relative error in computing the thermal conductivity*

The iterative procedure starts with computing the temperature due to the applied current (Joule temperature) using the thermal conductivity of the covalent Si. This is supplied to equation (2), thus, obtaining an initial estimate of the conductivity of the metallic phase. Following that a provisional value of the effective conductivity of the metallic layer and the covalent substrate is computed.

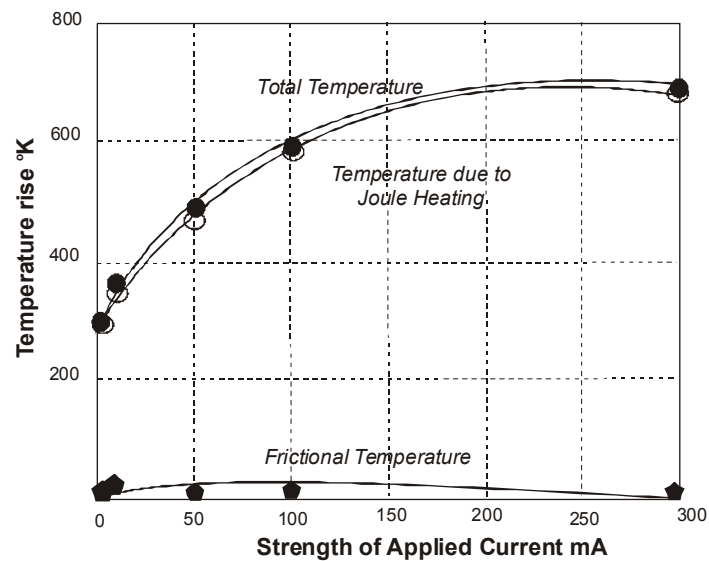

*Figure 5     Contribution of the two components of temperature (Joule Temperature and Frictional temperature) to the temperature rise during scratching Load= 5 g, sliding speed = 0.005m/s*

This is supplied to equation (9) to calculate the frictional contribution to the temperature rise in sliding. The obtained friction temperature is then added to the Joule temperature thus obtaining an updated value of the total temperature rise in the wafer surface. Such a value is used back in equation (2), the WFL law, to obtain yet another updated value of the thermal conductivity. The procedure is repeated until the difference between two successive values of the conductivity is smaller than a predetermined tolerance. For each set of measurements this procedure is applied to obtain the thermal conductivity of the metallic phase and the combination of the metallic phase and the covalent substrate. Figure 5 depicts the contribution of both components of the temperature Joule and frictional as a function of the strength of the applied current.

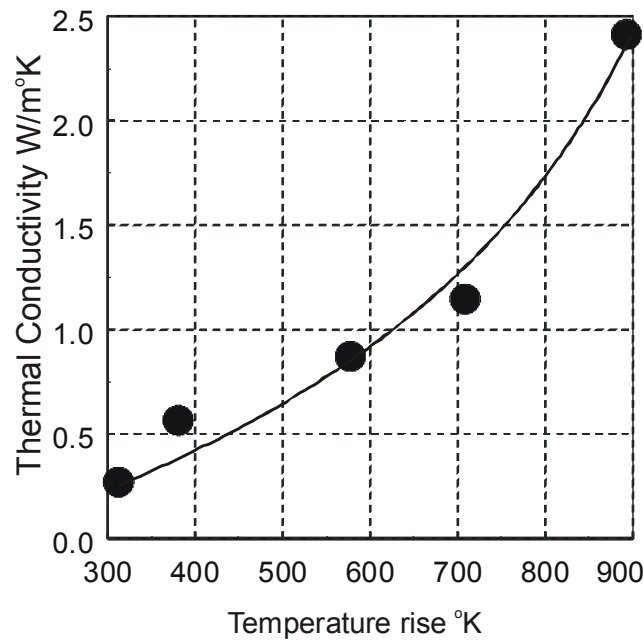

*Figure 6*   *Variation in the thermal conductivity of the metallic Si-Phase with temperature*

It is noted that the main contribution to the temperature rise of the surface comes from Joule heating. This is expected as the loads and the speeds used in this particular experiments are very slow (refer to table 1). Also since no provisions were made in the experiments to measure the coefficient of friction, the calculations had to rely on cited values in open literature. The estimates of the thermal conductivity obtained by this method are presented in figure 6. The values are plotted against the strength of the current and against the total temperature variations. While it is early at this stage of the research program, in light of the unavailability of exhaustive measurements, to comment on the behavior of the conductivity few remarks are in order:

    The values of the conductivity appear to be extremely lower than the conductivity of the covalent silicon (around 124 $Wm^{-1} K^{-1}$ for the later ).

The conductivity appears to have a dependency on the strength of the applied current as seen in figure 5-a. The values experience their highest value at extremely low currents.

Following that highest value a noticeable sharp decrease is noted. Past this point the conductivity increases with the increase in the strength of the current. Almost identical behavior is noted for the variation with respect to the temperature. This is in contrast to the behavior of the covalent silicon where a decrease in the conductivity is experienced with the increase in temperature.

It is curious that the computed values are in the same order of magnitude of the conductivity of some brittle metals and oxides. In particular the computed values are almost comparable to those of pure titanium, Sapphire ,and hematite to name a few.

It seems at this point of the research that the conductivity displays a dependency on loads, in fact unpublished preliminary computations for higher loads (0.1 N) yielded lower values of the conductivity. Such dependency is in line with the behavior of the conductivity for a strained solid [17,18]. Complete analysis of that case was considered out of the scope of this work. However, the remark of load dependency is worthy of further investigation.